\documentclass[prc,aps,showpacs,twocolumn,superscriptaddress,floatfix]{revtex4}

\usepackage{graphicx}
\usepackage{epsfig}
\usepackage{amsfonts}
\usepackage{amsmath,amssymb}
\usepackage{bm}

\begin{document}

\title{The information content of a new observable: the case of the nuclear neutron skin}

\author{P.--G. Reinhard}
\affiliation{Institut f\"ur Theoretische Physik II, Universit\"at
Erlangen-N\"urnberg, Staudtstrasse 7, D-91058 Erlangen, Germany}

\author{W. Nazarewicz}
\affiliation{Department of Physics \&
Astronomy, University of Tennessee, Knoxville, Tennessee 37996}
\affiliation{Physics Division, Oak Ridge National Laboratory, P.O. Box
2008, Oak Ridge, Tennessee 37831}
\affiliation{Institute of Theoretical Physics, University of Warsaw, ul. Ho\.za
69, 00-681 Warsaw, Poland}
\affiliation{School of Engineering and Science,
University of the West of Scotland,
Paisley  PA1 2BE, United Kingdom} 

\date{\today}

\begin{abstract}
We address two questions pertaining to the uniqueness and usefulness of a new observable: (i) Considering the current theoretical knowledge, what novel information does new measurement bring in?  (ii)
How can new data  reduce uncertainties of current theoretical models? 
We illustrate these points by studying the radius of the neutron distribution of a heavy nucleus, a quantity related to the equation of state for neutron matter that determines properties of nuclei and neutron stars. By systematically varying parameters of two theoretical models and studying the resulting confidence ellipsoid, we quantify the relationships between the neutron skin and various properties of finite nuclei and infinite nuclear matter. Using the  covariance analysis, we identify observables and pseudo-observables that correlate, and do not correlate,  with the   neutron skin. By adding the information on the neutron radius to the pool of observables determining the  energy functional, we show how precise experimental determination of the neutron radius in  $^{208}$Pb would reduce theoretical uncertainties on the neutron matter equation of state.
\end{abstract}

\maketitle

{\it Introduction.}
Nuclei communicate with us through a great variety of observables. Some are easy to measure, some take a considerable effort and experimental ingenuity.  Without any preconceived knowledge, all different observables are independent of each other and can usefully inform theory. On the other extreme, new data would be redundant if our theoretical model were perfect.  Reality lies in between. 
In this paper we show how
to assess the uniqueness and usefulness of an observable, i.e., its information content with respect to current theoretical models. We also quantify the meaning of  a correlation between different observables.

Let us consider a model characterized by a number of parameters $\mathbf{p}=(p_1,...,p_F)$ defining the {\it model space}. Those parameters may be, e.g.,  coupling constants of the effective Hamiltonian and  effective charges characterizing  operators in the assumed Hilbert space.  Calculated observables are functions of these parameters. Since the number of parameters is usually much smaller than the number of observables,  correlations  exist  between computed quantities. Moreover, since the model space has been optimized to a limited set of observables, there may also exist correlations between  parameters.

The paper is organized as follows. We first explain the
statistical approach used to estimate theoretical uncertainties and quantify correlations between calculated observables.   We next discuss the importance of the neutron skin measurement, briefly review related theoretical work, and describe  the model used and the set of fit-observables employed. 
The remainder of the paper contains results, conclusions, and  the outlook for the future.

{\it Regression analysis.}
Usually, most of the model space produces observables which are far from
reality. Therefore, one needs to confine the model space to a ``physically reasonable'' domain. That can be  achieved by a
least squares regression analysis. To this end, one selects a pool of fit-observables $\mathcal{O}$ 
which are used to calibrate $\mathbf{p}$.  
The optimum parameterization   $\mathbf{p}_0$ is  determined by a least squares fit with the global quality measure
\begin{equation}\label{chi2}
\chi^2(\mathbf{p})
=
\sum_{\mathcal{O}}
\left(
\frac{\mathcal{O}^\mathrm{(th)}(\mathbf{p})
     -
     \mathcal{O}^\mathrm{(exp)}}
    {\Delta\mathcal{O}}
\right)^2,
\end{equation}
where   ``th'' stands for the calculated values,
``exp'' for  experimental data, and $\Delta\mathcal{O}$
for  adopted errors. Having determined $\mathbf{p}_0$, an expectation value
of an observable $A$ can be computed at 
$A(\mathbf{p}_0)$.  
However, there remain uncertainties, originating both from the errors in 
fit-observables and from a limited
reliability of the model. To estimate the root-mean-square (rms) variation of $A$,
one needs to
define a physically reasonable 
domain around  $\mathbf{p}_0$.
Near the minimum,
the $\chi^2$-landscape 
is given by a
confidence  ellipsoid (see Sec.~9.8 of \cite{[Bra97a]}):
\begin{equation}\label{chi2a}
\chi^2(\mathbf{p})\!-\!\chi^2_\mathrm{0}
\approx 
\sum_{i,j=1}^F (p_i\!-\!p_{i,0})\mathcal{M}_{ij}(p_j\!-\!p_{j,0}),
\end{equation}
where  
\begin{equation}
\mathcal{M}_{ij}=
{\textstyle\frac{1}{2}}\partial_{p_i}\partial_{p_j}\chi^2|_{\mathbf{p}_0}.
\end{equation}
The physically reasonable domain $\mathbf{p}$ is defined as that
multitude of parameters around $\mathbf{p}_0$
that fall inside the covariance ellipsoid 
$\chi^2=\chi^2_0+1$,
i.e.,
\begin{equation}\label{confidence}
(\mathbf{p}-\mathbf{p}_0)\hat{\mathcal{M}}(\mathbf{p}-\mathbf{p}_0)
\leq 1.
\end{equation}
In terms of $\mathcal{M}_{ij}$, the covariance  between two observables $A$ and $B$ becomes:
\begin{equation}\label{cova}
\overline{\Delta A\,\Delta B}
=
\sum_{ij}\partial_{p_i}A(\hat{\mathcal{M}}^{-1})_{ij}\partial_{p_j}B
\quad.
\end{equation}
For $A$=$B$, Eq.~(\ref{cova})  gives variance
$\overline{\Delta^2 A}$ that defines an uncertainty of an observable. In addition, one can also establish the Pearson product-moment correlation coefficient between two
observables \cite{[Bra97a]}:
\begin{equation}
{c}_{AB}
=
\frac{|\overline{\Delta A\,\Delta B}|}
    {\sqrt{\overline{\Delta A^2}\;\overline{\Delta B^2}}}
\quad.
\label{correlator}
\end{equation}
A value ${c}_{AB}=1$ means fully correlated and ${c}_{AB}=0$ is
totally uncorrelated.

{\it Nuclear neutron skin.}
The covariance analysis is a standard statistical tool that  can be applied to {\it any} theoretical model 
that has been optimized to the data. In this work, we illustrate the general concept by considering one particular observable of fundamental importance for nuclear physics and astrophysics: the rms
radius of the neutron density distribution in a heavy nucleus, $r^{\rm  rms}_n=\langle r^2\rangle^{1/2}_n$. The size of $r^{\rm rms}_n$ is strongly correlated with  many properties characterizing neutron-rich matter found in neutron-rich nuclei \cite{[RISAC]} and in neutron stars \cite{[Hor01]}. 
The highly anticipated Lead Radius Experiment (PREX) at
Jefferson Laboratory  will use the parity-violating electroweak asymmetry in the elastic scattering of polarized electrons  to determine the neutron radius of $^{208}$Pb with a projected experimental precision of 1\%, in a model-independent fashion \cite{[Hor01a]}. Below, we apply the covariance analysis  to address questions pertaining to neutron-rich matter in general  and PREX experiment in particular:
(i) What  quantities that are experimentally accessible from finite nuclei correlate best, or do not correlate,  with neutron radius?
(ii) 
How robust are correlations between observables from finite nuclei and nuclear
matter properties (NMP)? 
(iii) 
To what extent would  precise data on the neutron radius in $^{208}$Pb enhance the predictive ability of theory?

A quantity that is related to $r^{\rm rms}_n$ is the neutron skin $r_{\rm skin} = r^{\rm rms}_n - r^{\rm rms}_p$ \cite{[Ton84],[Miz99]}.
The usefulness of neutron skin lies in its strong dependence on the isovector density $\rho_1=\rho_n-\rho_p$ and a much weaker dependence on the 
isoscalar, or total,  density $\rho=\rho_n+\rho_p$. 
A number of relationships, or  correlations, have been established between 
$r_{\rm skin}$ in heavy nuclei and 
various NMP and observables in finite nuclei (see \cite{[Ton84]} for an early discussion).
Those  include:  symmetry energy at the saturation point  $a_{\rm sym}(\rho_{\rm eq})$  \cite{[Ton84],[Rei99b],[Fur02],[Die03],[Che05],[Yos04],[War09],[Klu09]},
slope of bulk symmetry energy $a'_{\rm sym}=da_{\rm sym}/d\rho$ (proportional to the pressure difference between neutrons and protons) at $\rho_{\rm eq}$ 
\cite{[Rei99b],[Che05],[War09]} and at  $\rho$=0.1\,nucleons/fm$^{3}$ \cite{[Bal04]},
slope of binding energy of neutron matter $d(E/A)_n/d\rho_n$ at $\rho_n$=0.1\,neutrons/fm$^{3}$ \cite{[Bro00a],[Typ01],[Fur02],[Bod03],[Che05],[Yos04]} (proportional to the neutron pressure), 
the symmetry correction to the incompressibility $\Delta K$ \cite{[Fur02]},
low-energy electric dipole strength attributed to the Pygmy Dipole Resonance
(PDR) \cite{[Tso04],[Pie06],[Kli07]},
neutron form factor \cite{[Fur02]},
and $r_{\rm skin}$ in different nuclei \cite{[Sil05]}. 
It has also been found that
there are NMP which correlate poorly with
$r_{\rm skin}$: equilibrium nuclear matter binding energy and saturation density 
$\rho_{\rm eq}$ \cite{[Fur02]}, 
incompressibility $K$ \cite{[Fur02],[Yos04]}, and
enhancement factor of the Thomas-Reiche-Kuhn sum rule $\kappa_{\rm TRK}$ 
(related to the  isovector effective mass)  \cite{[Rei99b]}.

Some of the  previous theoretical papers dealing with neutron skin
correlations have explored the dependence between observables by
explicit variation of selected properties (e.g. symmetry energy)
within the given model, see, e.g.,  Refs.~\cite{[Ton84],[Klu09]}. The present
covariance analysis is the least biased and most exhausting way to
find out the correlations (\ref{cova}) between all conceivable
observables. There remain, however, what is called systematic errors
which are here hidden constraints and limitations of the given model.
Such systematic errors can only be determined by comparing different
models or sufficiently flexible variants of a model.  A comparison of
different models as in, e.g., Refs.~\cite{[Bro00a],[Typ01],[Fur02]} is thus an instructive complement.  But the use of
different models is not appropriate to quantitatively assess the
correlation between observables.  For that reason, our study is based
on the covariance analysis within the framework of one model.

{\it The Model.}
The theoretical approach employed in this study is the self-consistent mean field theory in  the  nuclear Density Functional
Theory (DFT) formulation \cite{[Ben03]}. At its heart  lies  the nuclear
energy  density functional (EDF) that is built from the nucleonic
intrinsic densities and - in a relativistic version - meson fields. The nuclear
DFT  framework has
been successful in describing a broad range of nuclear  properties,
including ground-state properties, excited states, particle decays, and
fission. Over the last few years, however, it has become evident  that the
standard functionals are too restrictive when one is aiming at the
detailed quantitative description and extrapolability.  Consequently, various strategies have been
devised to develop realistic EDF of spectroscopic quality
\cite{[Ber07a]}. Early attempts to employ statistical methods of
linear-regression and error analysis \cite{[Fri86]} have been revived
recently and been applied   
to determine the independence of EDF parameters, their errors, and the
errors of calculated observables.
\cite{[Ber05],[Kor08],[Toi08],[Klu09],[Sto09]}. The major
uncertainty in EDF lies in the  isovector channels that are poorly
constrained by experiment. In this context, neutron skin data are crucial.

The EDF used in this work is the Skyrme functional SV-min of
Ref.~\cite{[Klu09]}. It is characterized by $F$=14 coupling constants
(listed in Table~V therein).  The observables chosen to define
$\chi^2$ during optimization of SV-min embrace nuclear bulk properties
(binding energies, surface thicknesses, charge radii, spin-orbit
splittings, and pairing gaps) for selected semi-magic nuclei which are
proven to allow a reasonable DFT description.  For a list of chosen
observables, pseudo-observables, and adopted errors, see Tables~I-IX
of Ref.~\cite{[Klu09]}. NMP are not included in the fit data for
SV-min. This allows to count them as extrapolated observables in the
present correlation study. The parameter set $\mathbf{p}_0$ of SV-min
provides a very reasonable description of finite nuclei and nuclear
matter ($K$=222\,MeV, $a_{\rm sym}$=30.7\,MeV, effective nucleon mass
$m^*/m$=0.95).

A second EDF used in this study is the relativistic mean field.
We use it here in a traditional form in which Dirac nucleons
are coupled to finite-range meson fields: isoscalar scalar, vector,
isovector vector, and the Coulomb field and where the density
dependence is modeled only by non-linear couplings of the scalar field
\cite{[Rei89a],[Vre05]}.  This ``standard" model is too constrained in
the isovector channel and with respect to effective mass. It produces
too narrow covariance ellipsoids for our correlation analysis. Therefore we
augmented it by tensor couplings of vector fields \cite{[Rei89a]} and by
an isovector scalar field with mass 980 MeV, denoting the resulting functional
as RMF-$\delta$-t. We fit the model
parameters to the same pool of data as SV-min. Since the resulting NMP
of RMF-$\delta$-t ($K$=197\,MeV, $a_{\rm sym}$38\,MeV, $m^*/m$=0.59)
strongly deviate from the accepted values (as all traditional RMF
models) we use this model only to
discuss the robustness of our certain predictions and to illustrate
the model dependence of the statistical analysis.

{\it Results.}
In our study, we studied selected NMP and a number of observables related to isovector properties of finite nuclei such as neutron skins and radii, binding energy differences, and  dipole polarizability \cite{[Lip89]}. The latter one, the key quantity for static response, has been calculated within the RPA method:
\begin{equation}
\alpha_D
=
2\sum_{n\in\mathrm{RPA}}
(|\langle\Phi_n|\hat{D}|\Phi_0\rangle|^2/E_n),
\end{equation}
where $n$ runs over the excitation spectrum, $E_n$ is the excitation
energy of the RPA state $|\Phi_n\rangle$, and $\hat{D}$ the electric dipole operator (see Ref. \cite{[Rei99b]} for details of RPA calculations). We also investigated the energies of Giant Resonances: monopole (GMR), dipole (GDR), and quadrupole (GQR), and the low-energy dipole strength in neutron-rich nuclei: 
\begin{equation}\label{PDR}
B(E1; {\rm PDR})=\sum_{n, E_n<E_{\rm max}} B(E1,n)
\end{equation}
with $E_{\rm max}$=10\,MeV. The latter quantity is sometimes related to the
PDR strength \cite{[Tso04],[Pie06],[Kli07]}.

\begin{figure}
\includegraphics[angle=0,width=0.99\columnwidth]{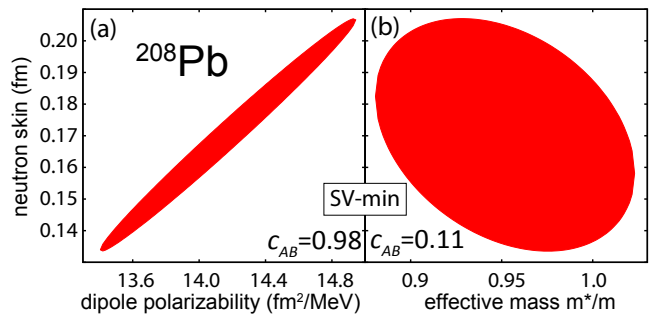}
\caption[T]{\label{fig:varellips-demo}
(Color online) The covariance ellipsoids  for two  pairs of observables as indicated.
The filled area shows the region of reasonable domain $\mathbf{p}$. 
Left: neutron skin and 
isovector dipole polarizability  in $^{208}$Pb. Right:
neutron skin in $^{208}$Pb and  effective nucleon mass $m^*/m$ in symmetric nuclear matter.
}
\end{figure}

\begin{figure}
\includegraphics[angle=0,width=0.95\columnwidth]{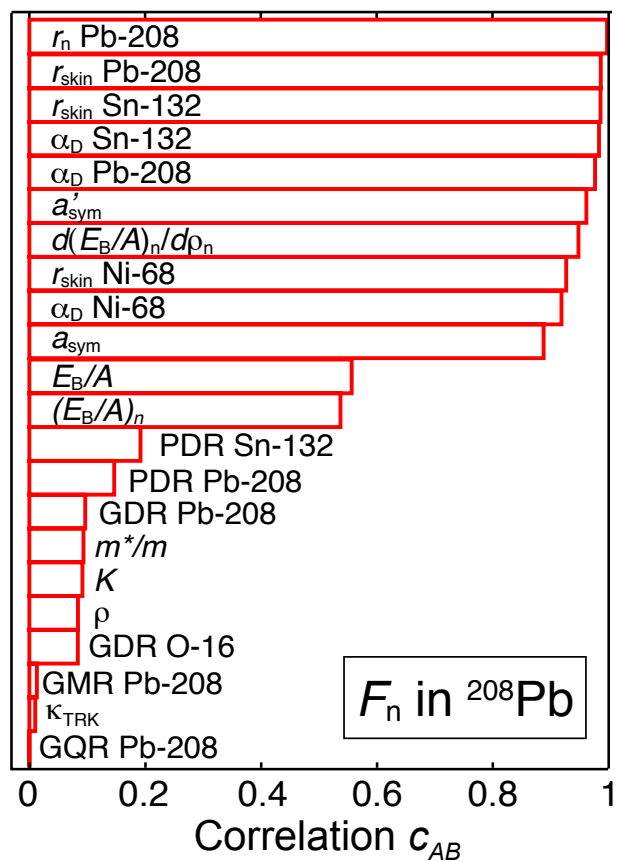}
\caption{\label{fig:aligns}
(Color online) Correlation (\ref{correlator}) of various observables with the neutron form factor
$F_n(q$=0.45\,fm$^{-1})$ in $^{208}$Pb. 
}
\end{figure}

Figure \ref{fig:varellips-demo} shows  covariance ellipsoids 
for two pairs of observables in $^{208}$Pb that nicely illustrate the cases  of  strong
correlation ($r_{\rm skin}$ and $\alpha_D$; $c_{AB}$=0.98) 
and weak correlation  ($r_{\rm skin}$ and $m^*/m$; $c_{AB}$=0.11).
Figure \ref{fig:aligns} shows correlations
with the point-neutron distribution form factor $F_n$ in $^{208}$Pb at $q$=0.45\,fm$^{-1}$ corresponding to PREX measurement. As expected, $F_n$  is strongly correlated with $r_{\rm skin}$, $r^{\rm  rms}_n$,  as well as with neutron skins in other neutron-rich nuclei. Almost equally strong is the correlation with the dipole polarizability. Not surprisingly, one can see excellent correlation of $F_n$ with NMP:  
$a_{\rm sym}$,  $a'_{\rm sym}$, and  $d(E/A)_n/d\rho_n$. 
All those quantities can thus be viewed as good indicators of isovector properties of nuclei.

The excellent correlation between the neutron skin  and  dipole polarizability is not surprising as $r_{\rm skin}\propto \alpha_D a_{\rm sym}$ \cite{[Sat06]}. 
The experimental value of $\alpha_D$ for $^{208}$Pb is 13.3$\pm$1.4\,fm$^2$/MeV
\cite{[Vey70]} while the value obtained by the Lorentz fit to the total
experimental photodisintegration cross section is 13.6\,fm$^2$/MeV  \cite{[Lip89]}. As seen in Fig.~\ref{fig:varellips-demo}, both values are  consistent with the  SV-min predictions for $r_{\rm skin}$. However, a 10\%   experimental uncertainty  due to  statistical
and photon-beam calibration errors makes it impossible to use the current best value of $\alpha_D$ as an
independent check  on neutron skin.

The nuclear and neutron matter binding energy seem poorly correlated with $F_n$, in accordance with Ref.~\cite{[Fur02]}. Our covariance analysis suggests the lack of correlation between $F_n$ (or neutron skin) and PDR strength; GMR, GDR, and GQR energies;
isoscalar and isovector effective mass, incompressibility, and saturation density (see also Refs.~\cite{[Fur02],[Yos04]}). Those quantities can thus be viewed as {\it poor isovector indicators}.

According to calculations, the degree of correlation with $F_n$ in $^{208}$Pb in general deteriorates with decreasing mass number
(see also discussion in \cite{[Ton84],[Yos04]}). This can be explained in terms of increased importance of shell effects  in  lighter nuclei. 
Shell effects are also responsible for the lack of correlation between
$r_{\rm skin}$ and PDR strength. The low-energy $E1$ strength is greatly impacted by the detailed single-particle structure around the Fermi level and thus varies
rapidly with EDF parameters. This reduces correlation with quantities which are weakly influenced by shell effects, e.g.,  NMP. 
The results of correlations for RMF-$\delta$-t are very similar to those from SV-min. In particular,  large correlations with neutron radii are predicted for $a_{\rm sym}$, and  $a'_{\rm sym}$ and small for $E/A$, $K$, and $m^*/m$.

\begin{figure}
\includegraphics[angle=0,width=0.95\columnwidth]{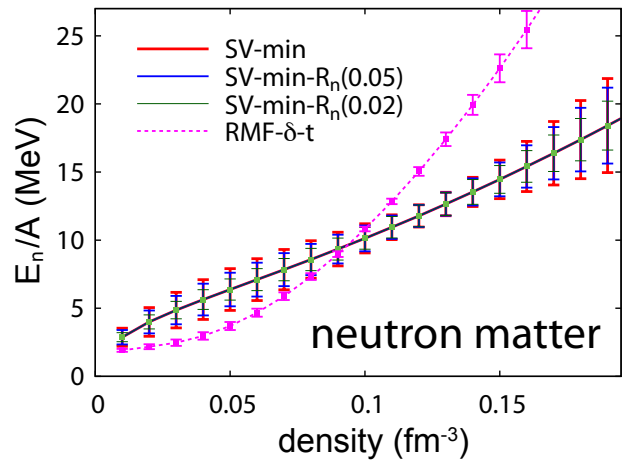}
\caption{\label{EOS}
(Color online)  Extrapolation errors for the neutron matter EOS predicted by  EDF SV-min  (obtained by a fit to the standard pool of data) and  SV-min-$R_n$ (obtained by adding to the data set the neutron
radius in $^{208}$Pb with an adopted error of 0.02\,fm and 0.05\,fm.
The neutron EOS predicted by RMF-$\delta$-t is also shown for comparison).
}
\end{figure}

To estimate the impact of precise experimental determination of neutron skin, we generated a new functional SV-min-$R_n$ by adding the value of neutron radius in $^{208}$Pb, 
$r^{\rm rms}_n$=5.61\,fm,
with an adopted error 0.02\,fm (0.4\% measurement)  and 0.05\,fm
(1\% measurement),
to the  set of fit observables. 
(The main difference between  SV-min and SV-min-$R_n$ is a slight reduction of 
isovector NMPs in the latter EDF: $a_{\rm sym}$ from 30.66\,MeV to 30.51\,MeV;
$a'_{\rm sym}$ from 92.73\,MeV\,fm$^3$ to 89.85 \,MeV\,fm$^3$; and $\kappa_{\rm TRK}$ from 
0.0765 to 0.057.)
Assuming a 0.4\% uncertainty in 
$r^{\rm rms}_n$, calculated  uncertainties on  isovector indicators shrink by about a factor of two. 
Figure~\ref{EOS} illustrates this tendency: it compares   
extrapolation errors for the neutron matter EOS in   EDF SV-min  and  SV-min-$R_n$. The impact of a 1\% measurement is much smaller, at least for the range of densities considered. 

We also carried out calculations with a new EDF obtained by a new fit where the neutron-rich nuclei have been given more weight (a factor 2 to 3 for
the three outermost neutron rich isotopes in most chains). 
The purpose of this exercise is to simulate the expected increased amount of data on neutron-rich nuclei.
While the  correlations seem to change very little, the extrapolation uncertainties in  neutron observables shrink by a factor of 1.5-2.0. For instance, with this new functional, the predicted neutron skin in $^{208}$Pb is $r_{\rm skin}$=0.191(0.024)\,fm, as compared to SV-min value of  $r_{\rm skin}$=0.170(0.037)\,fm. This exercise demonstrates that detailed conclusions of the statistical analysis depend on 
a chosen model  and a selected set of  fit observables. This point is also illustrated in Fig.~\ref{EOS}: the neutron matter EOS predicted  in SV-min is very different from that of RMF-$\delta$-t (which, as discussed earlier, yields
unphysical NMP). 

{\it Conclusions.}
In summary, we propose to use a statistical 
least-squares analysis to identify the impact of new observables, quantify correlations between predicted observables, and assess uncertainties of theoretical predictions. To illustrate the concept, we studied the the neutron radius of of $^{208}$Pb. By  means of covariance analysis
we identified a set of good isovector indicators
that correlate very well with the neutron form factor  of $^{208}$Pb.
These are: neutron skins and radii in neutron-rich nuclei,
dipole polarizability, and the nuclear matter properties such as symmetry energy and pressure. An indicator that is particularly attractive, as it can be measured in finite nuclei, is dipole polarizability. Unfortunately, the current best experimental value of $\alpha_D$ in $^{208}$Pb 
is not known precisely enough to 
offer  an independent check on neutron skin or to provide a quality constraint 
on EDF. 
We also demonstrate that 
nuclear and neutron matter binding energy, low-energy $E1$ strength, giant resonance energies,
isoscalar and isovector effective mass, incompressibility, and saturation density are  poor indicators of isovector properties, at least those related to $r_{\rm skin}$. 

We discussed the impact of PREX measurement on  theoretical  uncertainties for neutron-rich nuclei or neutron matter and concluded that it
will provide a  valuable constraint on the nuclear  energy functional that will reduce  theoretical error bars on the neutron-rich side.
While we have good reason to believe that our general conclusion about good and poor isovector indicators is robust, predictions for individual observables are obviously model dependent, as shown in Fig.~\ref{EOS}. This is an important point: even the best statistical analysis is not going to eliminate systematic errors due to incorrect theoretical assumptions. 

While our discussion is pertaining to the nuclear DFT, as the DFT is an obvious tool of choice to handle complex heavy nuclei and neutron skins, we believe that the methodology used in this work should be of interest to any theoretical framework that contains  parameters fine-tuned to experiment. Examples include fits of nucleon-nucleon forces to scattering and few-body data, adjustments of shell-model matrix elements, fits of coupling constants of symmetry-dictated Hamiltonians.

Discussions with M. Bender, P-H. Heenen, C. Horowitz,  and W. Satu{\l}a   are gratefully acknowledged.
This work was supported by BMBF under contract no. 06~ER~142D
and by the Office of Nuclear Physics,  U.S. Department of Energy under Contract
Nos. DE-FG02-96ER40963  and  DE-FC02-09ER41583.


\end{document}